\documentstyle[12pt,epsf,amssymb]{article}

\begin{document}

\title{On the Dynamical Stability of the Hovering Magnetic Top}
\author{S. Gov and S. Shtrikman\thanks{%
Also with the Department of Physics, University of California, San Diego, La
Jolla, 92093 CA, USA.} \\
%EndAName
The Department of Complex Systems, \\
Weizmann Institute of Science\\
Rehovot 76100, Israel \and H. Thomas \\
%EndAName
The Department of Physics and Astronomy\\
University of Basel\\
CH-4056 Basel, Switzerland}
\date{}
\maketitle

\begin{abstract}
In this paper we analyze the dynamic stability of the hovering magnetic top
from first principles without using any preliminary assumptions. We write
down the equations of motion for all six degrees of freedom and solve them
analytically around the equilibrium solution. Using this solution we then
find conditions which the height of the hovering top above the base, its
total mass, and its spinning speed have to satisfy for stable hovering.

The calculation presented in this paper can be used as a guide to the
analysis and synthesis of magnetic traps for neutral particles.
\end{abstract}

\section{Introduction}

\subsection{The hovering magnetic top.}

The hovering magnetic top is an ingenious device that hovers in mid-air
while spinning. It is marketed as a kit in the U.S.A. and Europe under the
trade name Levitron$^{\mbox{TM}}$ \hspace{0pt} \cite{levitron,patent} and in
Japan under the trade name U-CAS\cite{ucas}. The whole kit consists of three
main parts: A magnetized top which weighs about 18gr, a thin (lifting)
plastic plate and a magnetized square base plate (base). To operate the top
one should set it spinning on the plastic plate that covers the base. The
plastic plate is then raised slowly with the top until a point is reached in
which the top leaves the plate and spins in mid-air above the base for about
2 minutes. The hovering height of the top is approximately 3cm above the
surface of the base whose dimensions are about 10cm$\times $10cm$\times $%
2cm. The kit comes with extra brass and plastic fine tuning weights, as the
apparatus is very sensitive to the weight of the top. It also comes with two
wedges to balance the base horizontally.

\subsection{Qualitative Description.}

The physical principle underlying the operation of the hovering magnetic top
relies on the so-called `adiabatic approximation' \cite
{bergeman,Berry,bounds,coils}: As the top is launched, its magnetization
points antiparallel to the magnetization of the base in order to supply the
repulsive magnetic force which will act against the gravitational pull. As
the top hovers, it experiences lateral oscillations which are slow ( $\Omega
_{\mbox{lateral}}\simeq 1$Hz) compared to its precession ($\Omega _{\mbox{%
precession}}\sim 5$Hz). The latter, itself, is small compared to the top's
spin ($\Omega _{\mbox{spin}}\sim 25$Hz). Since $\Omega _{\mbox{spin}}\gg
\Omega _{\mbox{precession}}$ the top is considered `fast' and acts like a
classical spin. Furthermore, as $\Omega _{\mbox{precession}}\gg \Omega _{%
\mbox{lateral}}$ this spin may be considered as experiencing a {\em slowly}
rotating magnetic field. Under these circumstances the spin precesses around
the {\em local} direction of the field (adiabatic approximation) and, on the
average, its magnetization ${\bf \mu }$ points {\em antiparallel} to the
local magnetic field lines. In view of this discussion, the magnetic
interaction energy which is normally given by $-{\bf \mu }\cdot {\bf H}$ is
now given approximately by $\mu \left| {\bf H}\right| $. Thus, the overall
effective energy seen by the top is 
\begin{equation}
E_{\mbox{eff}}\simeq mgz+\mu \left| {\bf H}\right| .  \label{energy}
\end{equation}
By virtue of the adiabatic approximation, two of the three rotational
degrees of freedom are coupled to the transverse translational degrees of
freedom, and as a result the rotation of the axis of the top is already
incorporated in Eq.(\ref{energy}). Thus, under the adiabatic approximation,
the top may be considered as a {\em point-like} particle whose only degrees
of freedom are translational. The important point of this discussion is the
following: The energy expression written above {\em possesses a minimum} for
certain values of $\mu /m$. Thus, when the mass is properly tuned, the
apparatus acts as a trap, and stable hovering becomes possible.

As mentioned above, the adiabatic approximation holds whenever $\Omega _{%
\mbox{spin}}\gg \Omega _{\mbox{precession}}$ and $\Omega _{\mbox{precession}%
}\gg \Omega _{\mbox{lateral}}$. As $\Omega _{\mbox{precession}}$ is
inversely proportional to $\Omega _{\mbox{spin}}$, these two inequalities
can be satisfied simultaneously provided that the top is spun fast enough,
to get $\Omega _{\mbox{spin}}\gg \Omega _{\mbox{precession}}$, but not too
fast, for then $\Omega _{\mbox{precession}}>\Omega _{\mbox{lateral}}$!. The
reason for the {\em lower} bound on the spin is obvious: If the top is spun
too slowly, then $\Omega _{\mbox{spin}}\lesssim \Omega _{\mbox{precession}}$
and the top becomes unstable against {\em rotations}. The top will then flip
over and will be pulled quickly to the base. This is the instability that is
well known from classical top physics\cite{goldstein,landau}. The reason for
an {\em upper} bound on the spin is quite different: If the top is spun too
fast, the axis of the top becomes too rigid and cannot respond fast enough
to the changes of the direction of the magnetic field. It is then considered
as fixed in space and, according to Earnshaw's theorem\cite{earnshaw} ,
becomes unstable against {\em translations}.

\subsection{The purpose and structure of this paper.}

The major problem with the adiabatic approximation is that it cannot predict
exactly the allowed range of $\Omega _{\mbox{spin}}$ for which stable
hovering occurs. It simply gives us {\em estimates} for this range, and for
design purposes this may not be enough. The purpose of this paper is to give
a quantitative description of the physics of the hovering magnetic top while
it is in mid-air {\em without} using any preliminary assumptions (such as
the adiabatic approximation). To do this, we first expand the magnetic field
around the equilibrium point to second order in the spatial coordinates.
Using the gradients of this field we then find the force and torque on the
top and write down the equations of motion for all 6 degrees of freedom in
vectorial form\cite{simon}.{\em \ }Next, we solve these equations for the
stationary solution and for a small perturbation around this stationary
solution and arrive at a secular equation for the frequencies of the various
possible modes. The possible eigenmodes are either oscillatory
(corresponding to stable solution) or exponential (unstable solution). The
secular equation comes out linear in $\Omega _{\mbox{spin}}$, and it is not
difficult to write {\em analytic} expressions for the upper and lower bounds
of $\Omega _{\mbox{spin}}$.

The structure of this paper is as follows: In Sec.\ref{sec2} we first
describe our model and define our notations. Next, we derive the equations
of motion for the translational and rotational degrees of freedom, and
finally we solve these equations around the equilibrium position. In Sec.\ref
{sec3} we apply our results to the case of a disk-like top of radius $%
a=0.25R $ hovering above a circular current loop of radius $R$. We fix the
equilibrium hovering height and plot the various frequencies of the stable
modes as a function of $\Omega _{\mbox{spin}}$. We identify the origin of
the various modes and comment on the way these modes are coupled to produce
the minimum and maximum speeds at which the system becomes unstable. In
particular the connection of these results with the adiabatic approximation
will be discussed. Then we change the equilibrium height and note how the
allowed range of $\Omega _{\mbox{spin}}$ is affected. We shall show that
above a particular height, where according to the adiabatic approximation
hovering is not possible, hovering is {\em still} possible in a certain
range, and that, furthermore, the lower bound on $\Omega _{\mbox{spin}}\,$%
originates from a {\em new} mode coupling, not predicted by the adiabatic
approximation. In the last section we summarize our results and discuss the
possible uses of the derivation presented in this paper to the study of
magnetic traps for neutral particle.

\section{\label{sec2}Derivation and Solution of the Equations of Motion.}

\subsection{\label{sec2.1}Description of the model and notation.}

\begin{figure}[here]
\begin{tabbing}
%\hspace{1.5in} \=                           % horiz. space +
\epsfxsize=4.0in                        % width of figure
\epsffile{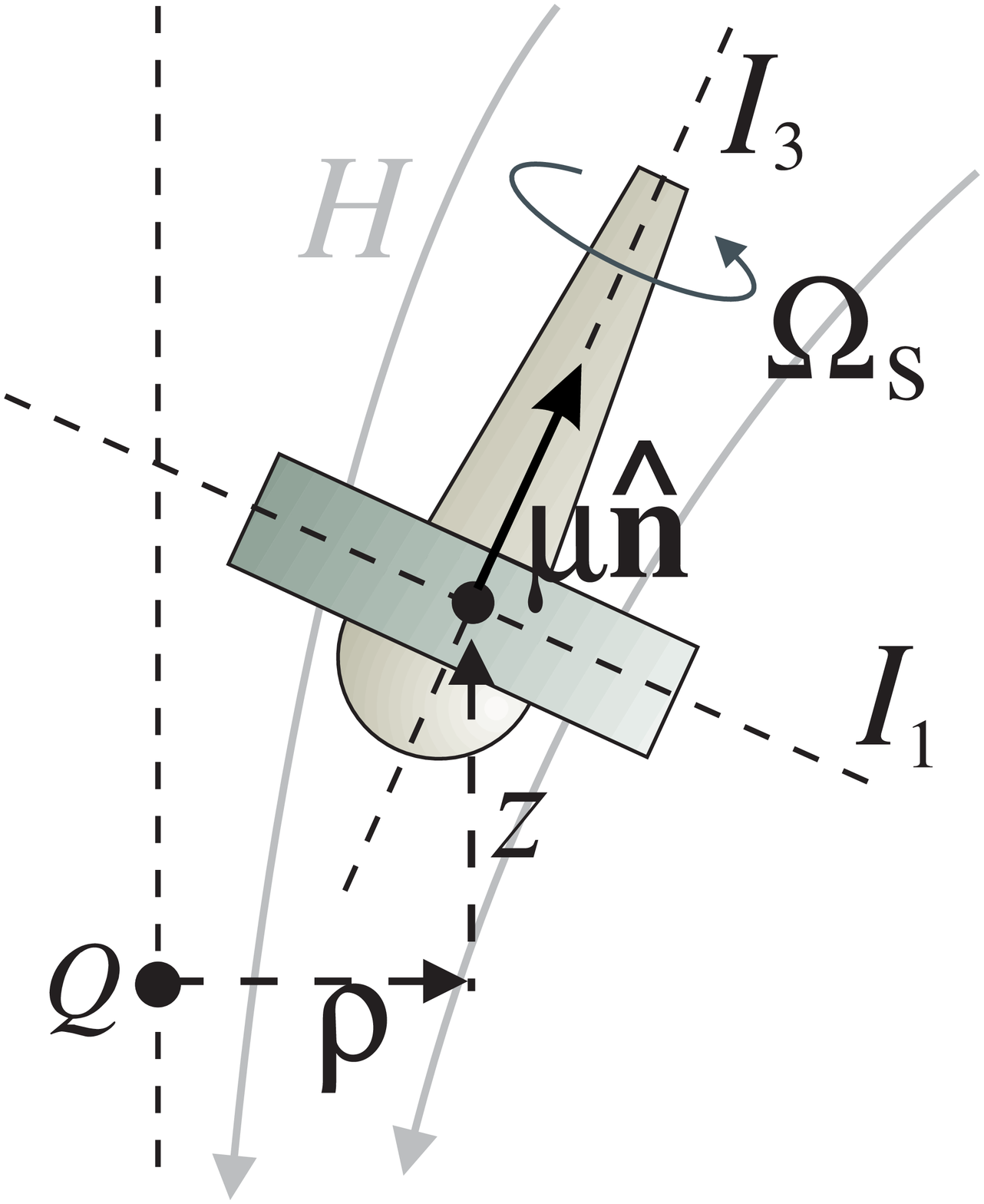} % read fig file
\end{tabbing}
\caption{The
hovering magnetic top near its equilibrium position (point Q) and notations.}
\label{fig1}
\end{figure}

In this paper we analyze the dynamics of the top while in mid-air as is
shown in Fig.\ref{fig1}. We consider a symmetric top with mass $m$ and a
magnetic moment $\mu $, rotating about its principal axis with angular speed 
$\Omega _{s}$. The magnetic moment of the top is assumed to be concentrated
at the center of gravity of the top and to point along the symmetry axis of
the top. The latter is denoted by the unit vector ${\bf \hat{n}}$. The
moment of inertia of the top about ${\bf \hat{n}}$ is $I_{3}\,$whereas the
secondary moment of inertia is $I_{1}$.

The spatial position of the top is given with respect to its equilibrium
position (point Q in the figure). Thus, $z$ is the vertical displacement of
the top and $\rho $ is its radial displacement. We denote by ${\bf \hat{z}}$
and ${\bf \hat{\rho}}$ the unit vectors in the vertical and radial
direction, respectively.
%==============================

%============================== 

Though this is not mandatory for our calculations, we assume for simplicity,
that the magnetic field possesses cylindrical symmetry around the vertical
axis. In addition, the direction of gravity is assumed to coincide with the
symmetry axis of the magnetic field and to point downward. We denote by $g$
the free fall acceleration.

\subsection{\label{sec2.2}The equations of motion.}

As the symmetry axis of the magnetic field coincides with the direction of
gravity, equilibrium is possible only along this axis. We therefore express
the magnetic field, to second order in $\rho $ and $z$, at the vicinity of
the equilibrium position in terms of its derivatives along the ${\bf \hat{z}}
$ direction {\em at} the equilibrium position. This is possible due to
cylindrical symmetry and to the fact that the Cartesian components of the
field are harmonics. The result is 
\begin{equation}
{\bf H}(\rho ,z){\bf =-}\frac{1}{2}\rho \left( H^{\prime }+H^{\prime \prime
}z\right) {\bf \hat{\rho}}+\left[ H+H^{\prime }z+\frac{1}{2}H^{\prime \prime
}\left( z^{2}-\frac{1}{2}\rho ^{2}\right) \right] {\bf \hat{z}}\mbox{.}
\label{eq5}
\end{equation}
Here $H$, $H^{\prime }$ and $H^{\prime \prime }$ are the vertical magnetic
field, its first and second derivatives along the $\hat{z}$ direction,
respectively, {\em at the equilibrium position, }i.e. at point Q.

The potential energy of the top is the sum of the magnetic interaction
energy of a dipole with a field plus a gravitational term, i.e. 
\[
E=-\mu {\bf \hat{n}}\cdot {\bf H}\left( \rho ,z\right) +mgz\mbox{.} 
\]
Consequently the force on the top is 
\begin{equation}
{\bf F}=-{\bf \nabla }E=\mu {\bf \nabla }\left( {\bf \hat{n}}\cdot {\bf H}%
\right) -mg{\bf \hat{z}}\mbox{,}  \label{eq1}
\end{equation}
whereas the torque is 
\begin{equation}
{\bf T}=\mu {\bf \hat{n}}\times {\bf H}\mbox{.}  \label{eq2}
\end{equation}

The equation for ${\bf r=}\rho {\bf \hat{\rho}+}z{\bf \hat{z}}$, the radius
vector of the center of mass of the top with respect to point Q, is given by 
\begin{equation}
m\frac{d^{2}{\bf r}}{dt^{2}}{\bf =}\mu {\bf \nabla (\hat{n}\cdot H)-}mg{\bf 
\hat{z}}\mbox{.}  \label{eq3}
\end{equation}

To write a vectorial equation of motion (see for example \cite{milne}) for
the angular momentum, ${\bf L}$ , we note that it has two components in
perpendicular directions. The first component is due to the rotation of the
top around the ${\bf \hat{n}}$ direction and is given simply by ${\bf L}%
_{n}=I_{3}\Omega _{s}{\bf \hat{n}}$. The second component of the angular
momentum is contributed by the {\em change} in the direction of the
principal axis from ${\bf \hat{n}}$ to ${\bf \hat{n}+}d{\bf \hat{n}}$.
Since, by definition, ${\bf \hat{n}}$ is a unit vector, it must point
perpendicular to $d{\bf \hat{n}}$. Thus, $\Omega _{\perp }=\left| d{\bf \hat{%
n}/}dt\right| $ is the angular velocity associated with the change of ${\bf 
\hat{n}}$. Since the direction of $\Omega _{\perp }$ must be perpendicular
to both $d{\bf \hat{n}}$ and ${\bf \hat{n}}$ we form the cross product ${\bf %
\Omega }_{\bot }={\bf \hat{n}\times }\left( d{\bf \hat{n}/}dt\right) $ which
incorporates both the correct value and the right direction. Multiplying $%
{\bf \Omega }_{\bot }$ by $I_{1}$ yields the second component of the orbital
angular momentum, ${\bf L}_{\perp }=I_{1}{\bf \hat{n}\times }\left( d{\bf 
\hat{n}/}dt\right) $. Thus, ${\bf L=L}_{n}+{\bf L}_{\perp }=$ $I_{3}\Omega
_{s}{\bf \hat{n}+}I_{1}{\bf \hat{n}\times (}d{\bf \hat{n}}/dt)$. Using this
expression together with Eq.(\ref{eq2}) for the torque we find that the
equation of motion for the angular momentum is 
\begin{equation}
\frac{d{\bf L}}{dt}=\frac{d}{dt}\left[ I_{3}\Omega _{s}{\bf \hat{n}+}I_{1}%
{\bf \hat{n}\times }\frac{d{\bf \hat{n}}}{dt}\right] =\mu {\bf \hat{n}}%
\times {\bf H}\mbox{.}  \label{eq4}
\end{equation}
Eqs.\ref{eq3},\ref{eq4} together form a coupled system of equations for all
6 degrees of freedom of the top. In these equations ${\bf r}$, ${\bf \hat{n}}
$ and $\Omega _{s}$ are the dynamical variables (a total of 6 degrees of
freedom), ${\bf H}$ is the magnetic field which itself is a function of $%
{\bf r}$, whereas $m$, $\mu $, $I_{1}$ and $I_{3}$ are external parameters.

\subsection{\label{sec2.3}Solution for the equations of motion}

The stationary solution of the problem is obvious: When the top is on the
symmetry axis with its principal axis parallel to the ${\bf \hat{z}}$ axis,
two forces act on the top. The first force is the downward gravitational
force $mg$ and the second one is the upward magnetic force supplied by the
external magnetic field, i.e., $\mu H^{\prime }$. Since these forces are
colinear, no torque is exerted. We therefore look for a solution of the form 
\begin{equation}
\begin{array}{c}
\rho (t)=z(t)=0 \\ 
{\bf \hat{n}}(t)={\bf \hat{z}} \\ 
\Omega _{s}(t)=\Omega _{s}=\mbox{Const.}
\end{array}
\label{eq6}
\end{equation}

Inserting this solution into Eqs.\ref{eq3},\ref{eq4} yields identities for 5
degrees of freedom. The equation for $z$, on the other hand, gives the
expected equilibrium condition 
\begin{equation}
\mu H^{\prime }=mg\mbox{.}  \label{eq7}
\end{equation}

To investigate the {\em stability} of the stationary solution, we now add
first order perturbations to the equations of motion. Thus, we make the
following substitutions 
\begin{equation}
\begin{array}{c}
z(t)\rightarrow 0+\delta z(t) \\ 
{\bf \vec{\rho}}(t)\rightarrow 0+\delta {\bf \vec{\rho}}(t) \\ 
\Omega _{s}(t)\rightarrow \Omega _{s}(t)+\delta \Omega (t) \\ 
{\bf \hat{n}}(t)\rightarrow {\bf \hat{z}}+\delta {\bf \hat{n}}(t)
\end{array}
\label{eq8}
\end{equation}
where, since ${\bf \hat{n}}$ is a unit vector, $\delta {\bf \hat{n}}$ must
be perpendicular to ${\bf \hat{z}}$. Substituting Eq.\ref{eq8} into Eqs.\ref
{eq3},\ref{eq4} and expanding to first order in the perturbations gives 
\begin{equation}
m\frac{d^{2}\delta {\bf \vec{\rho}}}{dt^{2}}=-\frac{1}{2}\mu H^{\prime }\delta {\bf \hat{n}} -\frac{1}{2}\mu H^{\prime \prime }\delta {\bf \vec{\rho}}\label{eq9.1}
\end{equation}
\begin{equation}
m\frac{d^{2}\delta z}{dt^{2}}=\mu H^{\prime \prime }\delta z  \label{eq9.2}
\end{equation}
\begin{equation}
\frac{d\delta \Omega }{dt}=0  \label{eq9.3}
\end{equation}
\begin{equation}
I_{3}\Omega _{s}\frac{d\delta {\bf \hat{n}}}{dt}+I_{1}{\bf \hat{z}}\times 
\frac{d^{2}\delta {\bf \hat{n}}}{dt^{2}}=-\mu H{\bf \hat{z}}\times \delta 
{\bf \hat{n}}-\frac{1}{2}\mu H^{\prime }{\bf \hat{z}}\times \delta {\bf \vec{%
\rho}}\mbox{.}  \label{eq9.4}
\end{equation}
We note that (at least to lowest order) the motions in the $\hat{z}$
direction and the rotation around the ${\bf \hat{n}}$ axis are {\em %
decoupled }from the other degrees of freedom. Furthermore, according to Eq.%
\ref{eq9.2} the motion in the $\hat{z}$-direction is stable provided that $%
H^{\prime \prime }<0$. We now focus our attention on the remaining four
degrees of freedom: Note that the right-hand side of Eq.\ref{eq9.1} contains
two terms: The first is the adiabatic term which tends to stabilize the top
against lateral translations by tilting the axis of the top. The second
term, which we call Earnshaw's term, tends to destabilize the top and to
take it away from the equilibrium position. Solving Eq.\ref{eq9.1} for $%
\delta {\bf \hat{n}}$ and substituting it into Eq.\ref{eq9.4} results in a
fourth order equation for the radius $\delta {\bf \vec{\rho}}$: 
\begin{equation}
\begin{array}{c}
2mI_{1}\hat{z}\times \delta {\bf \vec{\rho}}^{(4)}+2mI_{3}\Omega _{s}\delta 
{\bf \vec{\rho}}^{(3)}+ \\ 
+\mu \left[ I_{1}H^{\prime \prime }+2mH\right] \hat{z}\times \delta {\bf 
\vec{\rho}}^{(2)}+\mu I_{3}\Omega _{s}H^{\prime \prime }\delta {\bf \vec{\rho%
}}^{(1)} \\ 
-\mu ^{2}\left[ \frac{1}{2}\left( H^{\prime }\right) ^{2}-HH^{\prime \prime
}\right] \hat{z}\times \delta {\bf \vec{\rho}}=0
\end{array}
\mbox{.}  \label{eq10}
\end{equation}
The possible solutions of this equation are linear combinations of a steady
rotation of $\delta {\bf \vec{\rho}}$ around $\hat{z}$ at angular velocity $%
\omega $ (this is possible because of the cylindrical symmetry of the field;
otherwise, one should write two equations for the two components of $\delta 
{\bf \vec{\rho}}$). We thus set 
\begin{equation}
\frac{d\delta {\bf \vec{\rho}}}{dt}=\omega \hat{z}\times \delta {\bf \vec{%
\rho}}\mbox{.}  \label{eq10.1}
\end{equation}
Substitution of Eq.\ref{eq10.1} into Eq.\ref{eq10} results in the following
secular equation for the eigenfrequencies: 
\begin{equation}
\begin{array}{c}
2mI_{1}\omega ^{4}-2mI_{3}\Omega _{s}\omega ^{3} \\ 
-\mu \left[ I_{1}H^{\prime \prime }+2mH\right] \omega ^{2}+\mu I_{3}\Omega
_{s}H^{\prime \prime }\omega \\ 
-\mu ^{2}\left[ \frac{1}{2}\left( H^{\prime }\right) ^{2}-HH^{\prime \prime
}\right] =0\mbox{.}
\end{array}
\label{eq11}
\end{equation}

This fourth-order equation for $\omega $ has four real roots (or
eigenfrequencies) whenever the system is stable. Since we are looking for
the range of $\Omega _{s}$ for which the system is stable, we take another
point of view and express $\Omega _{s}$ in terms of $\omega $. The resulting
equation after using $mg=\mu H^{\prime }$ is

\begin{equation}
\Omega _{s}=\frac{2I_{1}H^{\prime }\omega ^{4}-g\left[ I_{1}H^{\prime
\prime }+2mH\right] \omega ^{2}-mg^{2}\left[ \frac{1}{2}H^{\prime }-\frac{%
HH^{\prime \prime }}{H^{\prime }}\right] }{2I_{3}H^{\prime }\omega
^{3}-gI_{3}H^{\prime \prime }\omega }\mbox{.}  \label{eq11.1}
\end{equation}

\section{\label{sec3}Application to a disk-like top above a circular current
loop.}

\begin{figure}[here]
\begin{tabbing}
%\hspace{1.5in} \=                           % horiz. space +
\epsfxsize=3.9288in      % width of figure
\epsffile{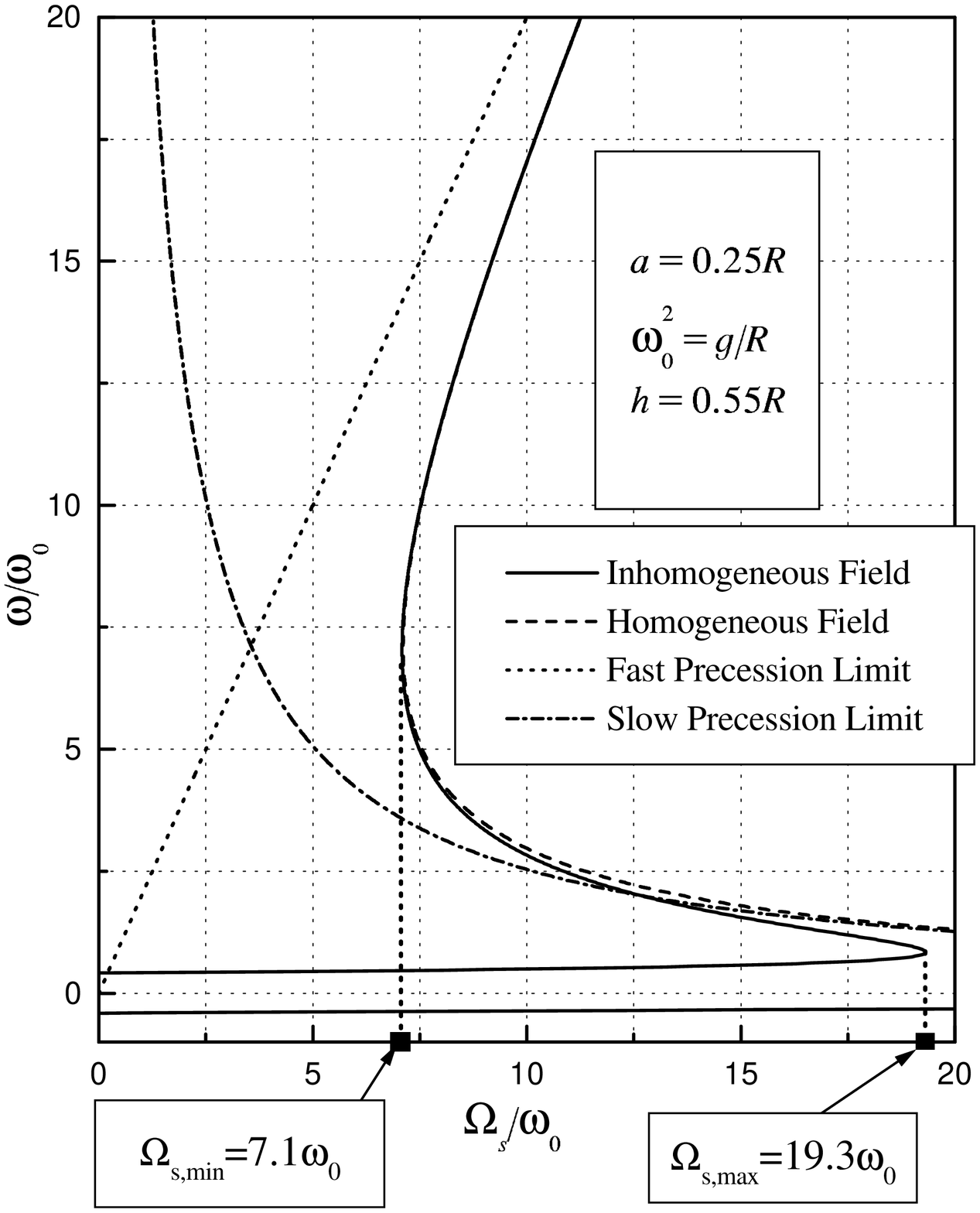} % read fig file
\end{tabbing}
\caption{Mode frequencies vs. spin for a disk-like top of radius $%
a$ hovering at a height $h$ above a circular current loop of radius $R$.
Solid line: inhomogeneous field, dashed line: homogeneous field, dotted
line: fast precession limit, dashed-dotted line: slow precession limit.}
\label{fig2}
\end{figure}

As an example we now apply Eq.\ref{eq11.1} to the case of a disk-like top of
radius $a$. Consequently, $I_{3}=ma^{2}/2$ and $I_{1}=ma^{2}/4$. The source
of the magnetic field is taken as a horizontal current loop (or
alternatively, a vertically uniformly magnetized thin disk) of radius $R$.
The vertical magnetic field and its derivatives along the axis at a height $%
h $ above the loop are therefore given by 
\begin{equation}
\begin{array}{c}
H=-H_{0}\left[ 1+\left( h/R\right) ^{2}\right] ^{-3/2} \\ 
H^{\prime }=\frac{3H_{0}}{R}\left( h/R\right) \left[ 1+\left( h/R\right)
^{2}\right] ^{-5/2} \\ 
H^{\prime \prime }=\frac{3H_{0}}{R^{2}}\left[ 1-4\left( h/R\right)
^{2}\right] \left[ 1+\left( h/R\right) ^{2}\right] ^{-7/2}\mbox{.}
\end{array}
\label{eq5.1}
\end{equation}
Note that the magnetic field was chosen to point downward. Thus, in order to
get an upward repulsive magnetic force, the magnetic moment should point
upward. The sign convention is chosen such that $h$ and $\mu $ are positive, 
$H$ is negative and, consequently, $H^{\prime }$ is positive. For stability
in the $z$ direction we require $H^{\prime \prime }<0$. This occurs,
according to Eq.\ref{eq5.1}, as long as $h>h_{\min }=0.5R,$ and sets a lower
bound for the height at which stable hovering is possible. Taking $h=0.55R$
and $a=0.25R$ (these are approximately the parameters for the Levitron)
inside Eq.\ref{eq11.1} and plotting $\Omega _{s}$ versus $\omega $ yields
the solid line shown in Fig.\ref{fig2}. Note that both $\Omega _{s}$ and $%
\omega $ are normalized to $\omega _{0}=\sqrt{g/R}$.
%==============================
%============================== 
From this figure we learn that (for $h=0.55R$ and $a=0.25R$) whenever 
\[
7.1\sqrt{g/R}<\Omega _{s}<19.3\sqrt{g/R} 
\]
there are four real solutions $\omega $ which correspond to four stable
modes. The frequency of the fastest mode goes asymptotically to the dotted
line $\Omega _{s}=\left( I_{1}/I_{3}\right) \omega $ as $\Omega _{s}\gg
\omega _{0}$. This is nothing but the `fast precession' rotational mode
encountered in classical top physics\cite{goldstein}. The frequency of the
next fastest mode goes roughly like $\omega \simeq \mu H/I_{3}\Omega _{s}$
(compare it to the dash-dash-dotted line). This is the well known `slow
precession' rotational mode of a classical top\cite{goldstein}. The coupling
between these two modes produces the {\em minimum} speed for stability $%
\Omega _{s,\min }=7.1\sqrt{g/R}$. For comparison purposes we have also
included in our plot (dashed line) the resultant mode frequencies when we
set $H^{\prime }$ and $H^{\prime \prime }$ $=0$ inside Eq.(\ref{eq11}). This
is equivalent to solving the problem of a magnetized top, in a {\em %
homogeneous} magnetic field. In fact, these mode frequencies may be obtained
as the roots of $\Omega _{s}=\left( I_{1}/I_{3}\right) \omega +\mu
H/I_{3}\omega $. The minimum value of this expression occurs for $\omega
^{2}=\mu H/I_{1}$ and is given by $\Omega _{s,\min }=\sqrt{4\mu
HI_{1}/I_{3}^{2}}$ , which is the minimum speed of a classical top in a
homogeneous field. By comparing the mode frequencies for the homogeneous
field to the mode frequencies for the inhomogeneous field we see that (as
far as the minimum speed is concerned) the minimum speed in both graphs is
almost the same. The two slowest modes in the solid-line plot are the two
vibrational modes of the top. It is clearly seen that one of them is
strongly coupled to the slow precession mode. This coupling is responsible
for producing the {\em maximum} speed for stability $\Omega _{s,\mbox{max}%
}=19.3\sqrt{g/R}$, as was already predicted by the adiabatic approximation.
But unlike in the adiabatic approximation, we now have a way to find
analytically both the minimum and maximum speed in a single stroke.

\begin{figure}[here]
\begin{tabbing}
%\hspace{1.5in} \=                           % horiz. space +
\epsfxsize=3.8795in    % width of figure
\epsffile{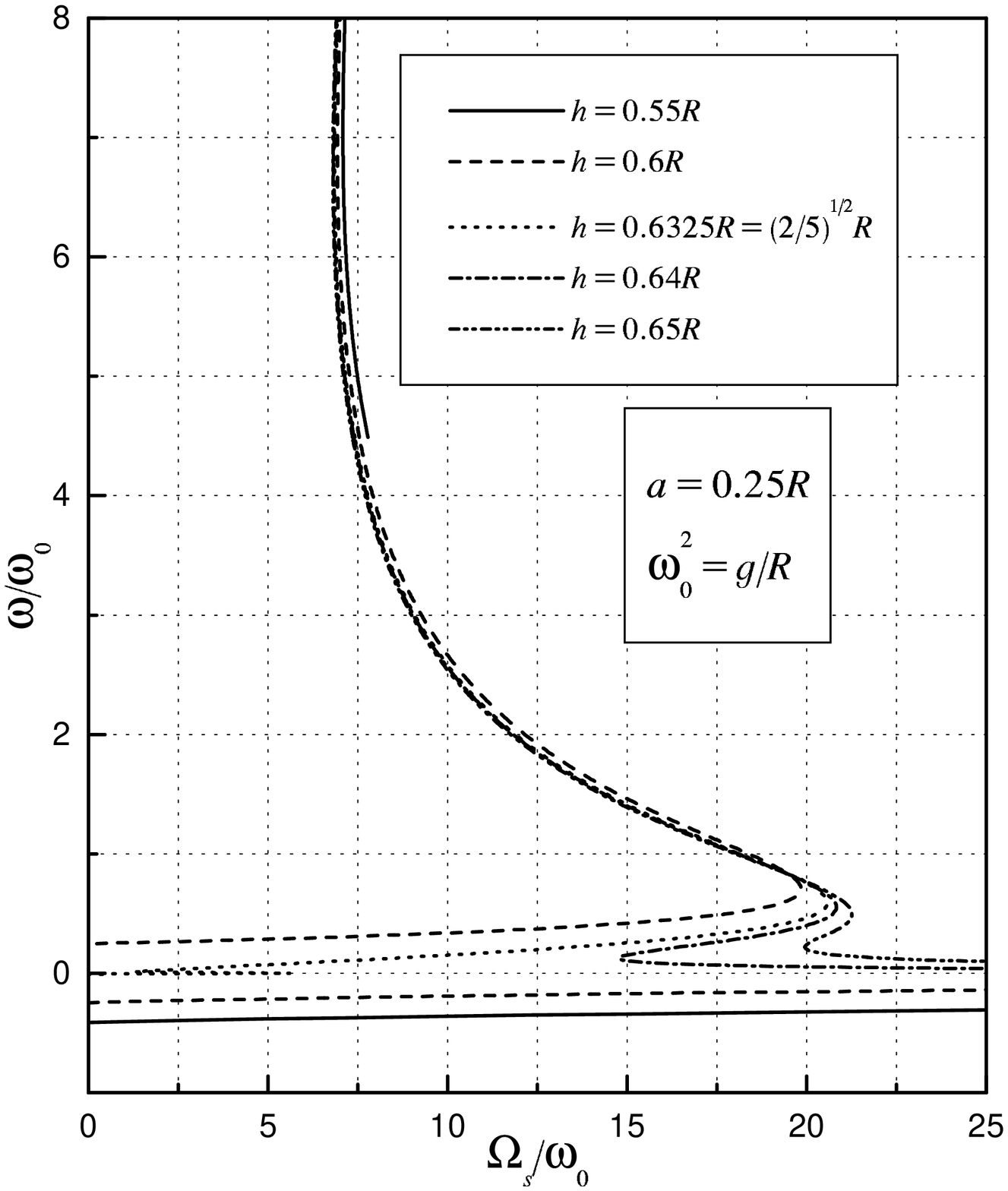} % read fig file
\end{tabbing}
\caption{Mode frequencies
vs. spin for a disk-like top of radius $a$ hovering at a height $h=0.55R,$ $%
0.6R$, $0.6325R,$ $0.64R$ and $0.65R$ above a circular current loop of
radius $R$.}
\label{fig3}
\end{figure}

\begin{figure}[here]
\begin{tabbing}
%\hspace{1.5in} \=                           % horiz. space +
\epsfxsize=3.87in   % width of figure
\epsffile{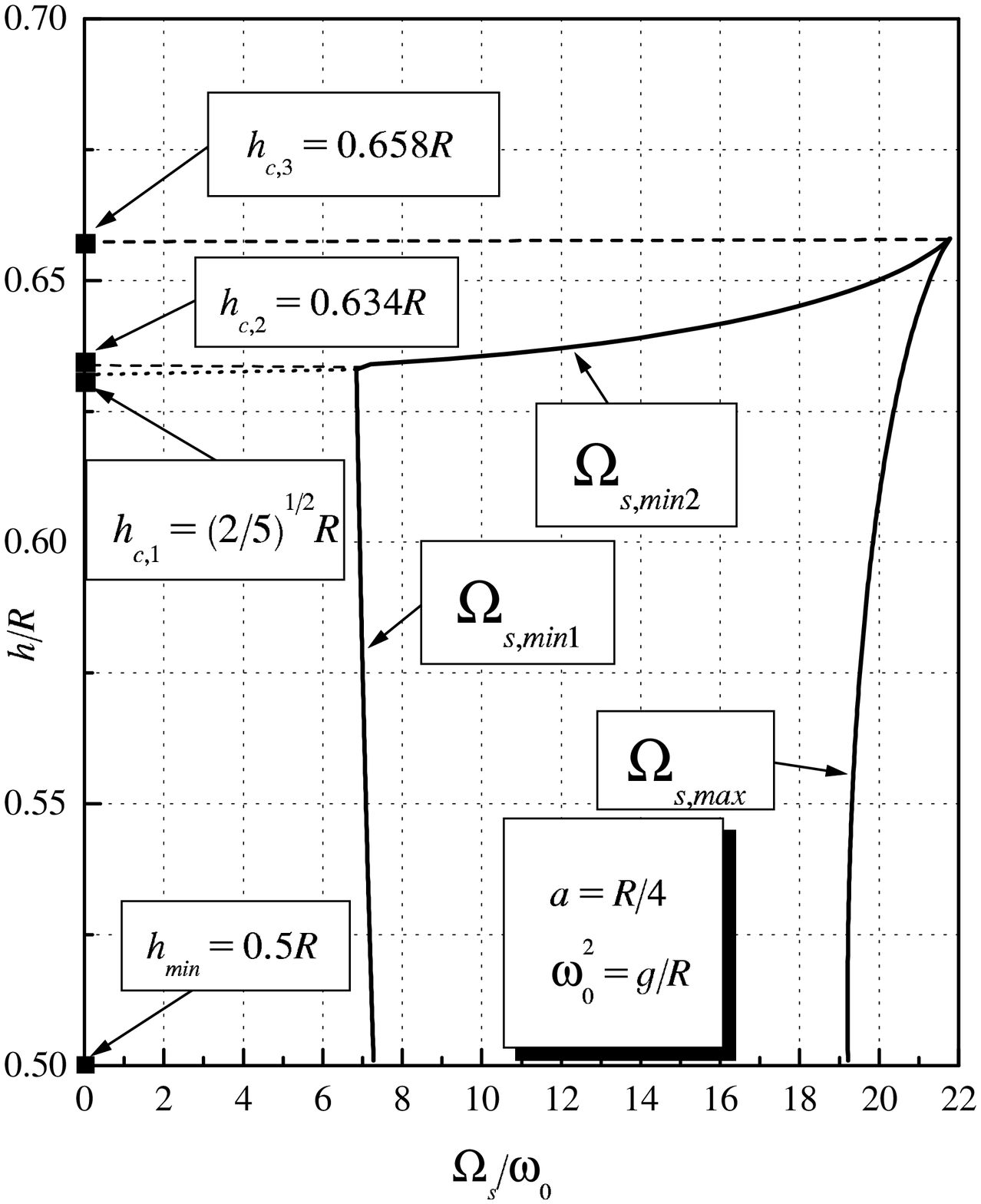} % read fig file
\end{tabbing}
\caption{Boundaries of the stability
region in the $\Omega _{s}$-$h$ plane.}
\label{fig4}
\end{figure}

Our next step is to study how the allowed range of $\Omega _{s}$ depends on
the equilibrium height $h$. In Fig.\ref{fig3} we have plotted the mode
frequencies versus $\Omega _{s}$ for $h=0.55R,$ $0.6R,$ $0.6325R$ $,$ $0.64R$
and $0.65R$.
%==============================
%============================== 
In Fig.\ref
{fig4} we have plotted the minimum and maximum speeds for each equilibrium
height starting at $h=0.5R.$ Recall that stability along the ${\bf \hat{z}}$
direction requires that $h>0.5R$. Thus, the two curves plotted in Fig.\ref
{fig4} together with the $h=0.5R$ line define a closed region in the $\Omega
_{s}$-$h$ plane. Each point {\em inside} this region corresponds to a stable
hovering solution whereas each point outside this region belongs to an
unstable solution. From this figure we may deduce that the range of heights
for which stable hovering is possible is given by 
\[
0.5R\leq h\leq 0.658R 
\]
The figure also shows that as $h$ increases above its minimum value $h_{%
\mbox{min}}=0.5R$, the minimum allowed speed $\Omega _{s,\mbox{min1}}$
decreases slightly, whereas the maximum allowed speed $\Omega _{s,\mbox{max}%
} $ increases. As $h$ increases further, the coupling between the two
vibrational modes becomes stronger, and when $h$ exceeds a first critical
value $h_{c,1}=\sqrt{2/5}R$, this coupling gives rise to the appearance of a 
{\em new} minimum $\Omega _{s,\mbox{min2}}$ of $\Omega _{s}\left( \omega
\right) $, which increases steeply with increasing $h$. As the height
exceeds the slightly larger value $h_{c,2}=0.634R$, this new minimum $\Omega
_{s,\mbox{min2}}$ becomes {\em higher} than the minimum speed $\Omega _{s,%
\mbox{min1}}$ determined by the coupling between the rotational modes, and
thus limits the stability of the top.

At $h=h_{c,3}=0.658R$, the minimum speed $\Omega _{s,\mbox{min2}}(h)$
crosses the maximum speed $\Omega _{s,\mbox{max}}(h)$, such that stable
hovering is not possible for $h>h_{c,3}$ and $\Omega _{s}>\Omega _{s,\mbox{%
min2}}(h_{c,3})$.

\begin{figure}[here]
\begin{tabbing}
%\hspace{1.5in} \=                           % horiz. space +
\epsfxsize=4.15in   % width of figure
\epsffile{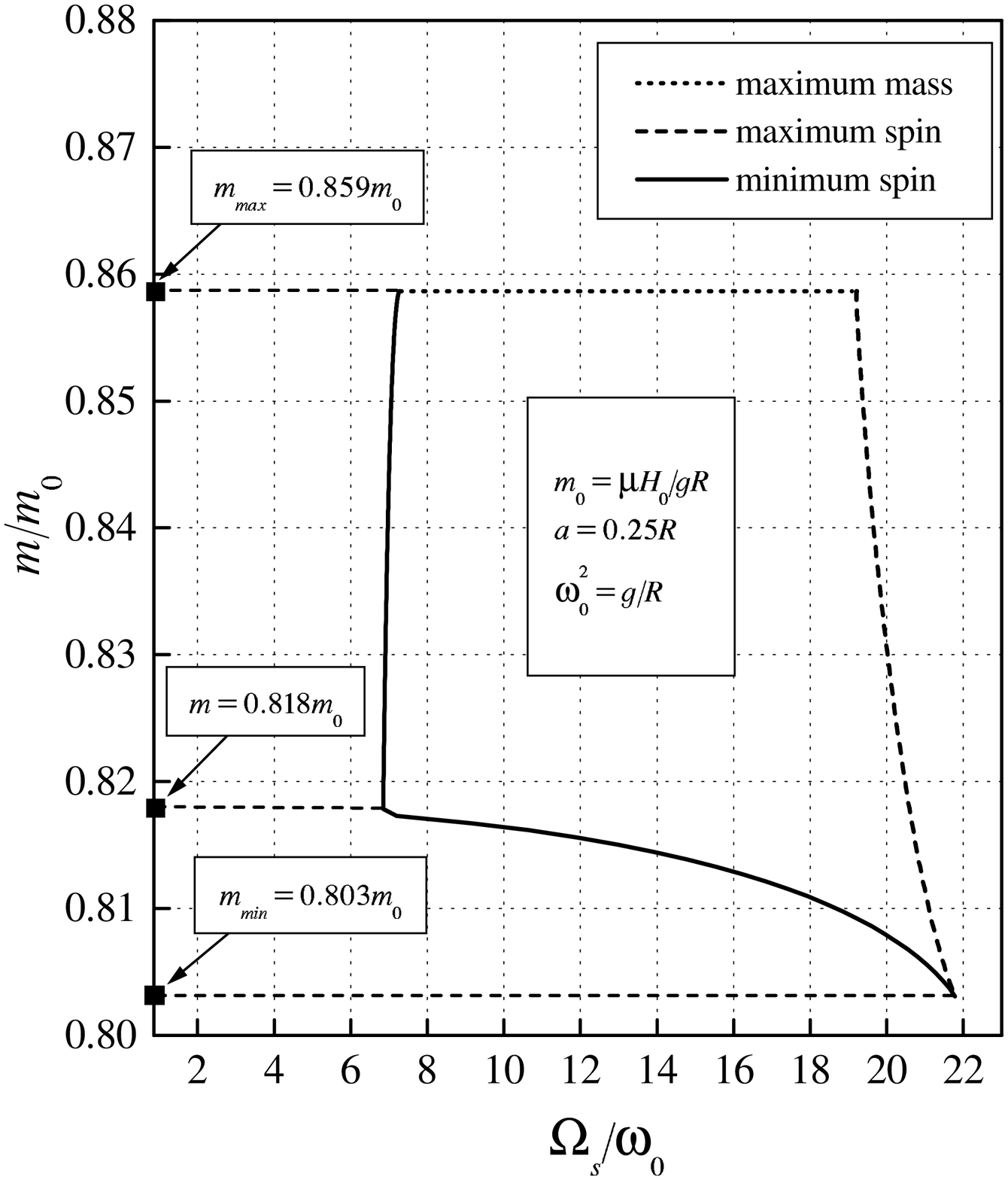} % read fig file
\end{tabbing}
\caption{Boundaries of the stability
region in the $\Omega _{s}$-$h$ plane.}
\label{fig5}
\end{figure}

It is important to note that the adiabatic approximation does not predict
this new coupling. According to the adiabatic approximation the system will
be stable against lateral translations whenever the curvature of the
effective energy (given in Eq.\ref{energy}) along the ${\bf \hat{\rho}}$
direction is positive. It can be shown that this is satisfied whenever $h<%
\sqrt{2/5}R.$ Thus, the adiabatic approximation predicts that hovering above 
$h=\sqrt{2/5}R$ is not possible. In the present calculation we find,
however, that the top is also stable above this height. This is due to the
splitting of the vibration degenracy by the coupling to the precessional
mode present when the spin is finite. This has already been pointed out by
M. V. Berry\cite{Berry} who treated it in terms the phenomenon of
geometrical magnetism.
%==============================
%============================== 
Since each
equilibrium height corresponds to a different repulsive magnetic force (and
hence to a different {\em mass}) we can, alternatively, specify the allowed
range of {\em mass} required for stable hovering by using Eq.\ref{eq7} and
Eq.\ref{eq5.1} (note that the maximum height corresponds to the lightest
mass and the minimum height corresponds to the heaviest mass). The results
are plotted in Fig. \ref{fig5}.
%==============================
%============================== 
From the
figure we learn that, for $a=0.25R$, the mass of the top must be such that 
\[
0.803m_{0}\leq m\leq 0.859m_{0} 
\]
where 
\[
m_{0}\equiv \frac{\mu H_{0}}{gR} 
\]
This corresponds to a mass tolerance of 
\begin{equation}
\frac{\Delta m}{m}\equiv 100\frac{m_{\mbox{max}}-m_{\mbox{min}}}{m_{\mbox{%
max}}}\simeq 7\%\mbox{.}  \label{eq.tol}
\end{equation}
Experimentally the tolerance is only about $1\%$ which seems like a large
discrepancy. Note, however, that the {\em lower} mass region (of less than $%
0.818m_{0}$) is difficult to access. Also, the top retains a finite kinetic
energy when launched into the trap which, as has been discussed by us\cite
{tolerance}, further decreases the mass tolerance in practice. Last but not
least, the theoretical mass tolerance, given in Eq.(\ref{eq.tol}) decreases
drastically with the tilt of the base, and goes to zero for a tilt of about $%
0.45$ degrees. This, in turn, makes it difficult to realize, in practice,
the above theoretical mass tolerance, even with the leveling wedges supplied
with the Levitron.

Note also that the temperature coefficient of the magnetization of the
ceramic magnet from which the Levitron is made is about $0.2\%$ per degree
Celsius. As the magnetic force varies with the {\em square} of the
magnetization, temperature changes may easily `throw' the mass out of range.

To overcome these problems the kit comes with a set of light washers to tune
the mass properly.

\section{\label{sec4}Summary.}

In this paper we have analyzed the hovering magnetic top while it is in
mid-air {\em without} using any preliminary assumptions. To do this we
expanded the magnetic field around the equilibrium point to second order in
the spatial coordinates. Using the gradients of this field we then found the
force and torque on the top and wrote down vectorial equations of motion for
all 6 degrees of freedom.{\em \ }Next, we solved these equations
analytically for the stationary solution and for a small perturbation around
this stationary solution and arrived at a secular equation for the
frequencies of the various possible modes. We then applied the solution to
the case of a disk-like top hovering above a circular current loop, and were
able to predict both the minimum and maximum allowed speeds for stable
hovering to occur.

Although the numerical results we have presented in this paper refer to the
case of a {\em disk-like} top ($I_{3}/I_{1}=2$), our theory is valid for the
whole range of the anisotropy parameter $I_{3}/I_{1}$ of the symmetric top,
and is therefore more general than the model studied in Ref.\cite{simon} who
approximated the top by a classical spin. We recover their results by
setting $I_{1}=0$ in our calculations\cite{ione=zero}. Also, our analysis
yields naturally the {\em minimum} spin for stability, which is zero for the 
$I_{1}=0$ case, and thus is put in `by hand' in the previous \cite{simon}
treatment. The determination of the {\em maximum} spin, as a function of the
anisotropy parameter $I_{3}/I_{1}$, is of particular interest for {\em %
cigar-like} tops ($I_{3}/I_{1}<1$), as for such tops the {\em minimum} speed
behaves drastically different than for {\em disk-like} tops ($%
2>I_{3}/I_{1}>1 $), as we have shown elsewhere\cite{sync1,tobe}.

An interesting corollary of our analysis is a limitation on the radius of
the {\em disk-like} top. One would have expected that increasing the radius
of the top will facilitate the operation of the Levitron as the moment of
inertia increases, thus reducing the {\em minimum} speed. However, the {\em %
maximum} speed decreases {\em faster} when the radius increases, with the
result that 
\[
\frac{\Omega _{s,\mbox{max}}}{\Omega _{s,\mbox{min}}}\simeq 0.66\frac{R}{r}%
. 
\]
Thus, the top cannot exceed two thirds of the base. Another detrimental
effect of increasing the radius is the reduction of the hovering time\cite
{friction} due to the increased effect of air friction. Moreover, air
friction also modifies the stability analysis: So far we carried out a
detailed study of the effect of friction on stability only for the $I_{1}=0$
model. In this case we found\cite{tobe} that with friction the top is {\em %
always }unstable even if the friction is infinitesimaly small and even if we
invoke translational viscosity only. We expect this behavior to occur also
for $I_{1}\neq 0$ indicating that spin traps are different in character from
potential traps in which friction `increases' stability.

We have disregarded here the angular momentum carried by the electrons
responsible for the ferromagnetic moment of the top. Although this is very
small compared to the orbital angular momentum of the Levitron, the two
angular momenta may become comparable for sufficiently small tops, which
would result in a left-right asymmetry. Furthermore, the possibility of
levitation based only on the {\em electronic} angular momentum arises. This
will be discussed elsewhere\cite{tobe}.

Also, our treatment is completely classical. As size decreases,
quantum-mechanical effects may become important. In particular, a
sufficiently small particle, for example an atom, will only have a finite
life-time in the trap due to quantum-mechanical effects, which will be
considered elsewhere\cite{tobe}.


\begin{thebibliography}{99}
\bibitem{simon}  The same problem can also be treated by approximating the
top by a classical spin. This is done by M. D. Simon, L. O. Heflinger and S.
L. Ridgway, {\em Am. J. Phys. }{\bf 65} (4), 286-292 (1997).
\end{thebibliography}
\end{document}